\documentclass[preprint,nofootinbib]{revtex4}%
\usepackage{amssymb}
\usepackage{amsfonts}
\usepackage{amsmath}
\usepackage{graphicx, tikz}
\usepackage{color}%
\usepackage{hyperref}
\usepackage{svg}
\providecommand{\U}[1]{\protect\rule{.1in}{.1in}}
\hypersetup{colorlinks,linkcolor={blue},citecolor={blue},urlcolor={black}} 
\providecommand{\U}[1]{\protect\rule{.1in}{.1in}}
\definecolor{blue}{rgb}{0,0,1}

\definecolor{red}{rgb}{1,0,0}

\begin{document}
\title{Black holes and phase transitions in Conformal Gravity}
\author{Luis Guajardo$^1$, Luis Gutiérrez Frez$^2$, Seyed Naseh Sajadi$^{3,4}$,  Julio Oliva$^5$}
\affiliation{$^1$ Instituto de Matem\'atica, F\'isica y Estad\'istica, Facultad de Ingenier\'ia y Negocios,
Universidad de las Am\'ericas, Sede Concepci\'on,
Avenida Jorge Alessandri Rodríguez 1160, 4090940, Chile\\
$^2$ Instituto de Ciencias Físicas y Matemáticas, Universidad Austral de Chile, Valdivia, Chile\\
$^3$ Department of Mathematics and Computer Science, Faculty of Science, Chulalongkorn University, Bangkok 10330, Thailand\\
$^4$ Strong Gravity Group, Department of Physics, Faculty of Science, Silpakorn University, Nakhon Pathom 73000, Thailand\\
$^5$ Departamento de Física, Universidad de Concepción, Casilla, Concepcion, 160-C Chile}
\begin{abstract}
We explore the thermodynamic properties of static black holes in conformal gravities in arbitrary, even dimensions. In the spherically symmetric case, up to local Weyl rescalings, these spacetimes are static and unique, due to a Birkhoff's theorem that applies for all conformal gravities whose Lagrangians are complete contractions of $k$ Weyl tensors in dimension $D=2k$. The solutions can be chosen to be asymptotically locally (A)dS or flat, and the horizon can be extended to be a surface of arbitrary constant curvature, $\gamma$. We focus on the asymptotically AdS case, for which the cosmological constant at infinity is an integration constant, and compute the mass of the spacetime in terms of the diverse hair parameters which have a purely gravitational origin. We show that the first law of thermodynamics is satisfied, and study phase transitions in the canonical ensemble, showing the existence of zeroth-order black-hole/black-hole phase transitions for arbitrary dimension. 
\end{abstract}
\maketitle

\section{Introduction}

In four dimensions, the conformal algebra $\mathfrak{so}(4,2)$ emerges as the largest bosonic, spacetime
symmetry algebra that a non-trivial field theory can have \cite{Coleman:1967ad}. Field
theories with this symmetry group in Minkowski spacetime can be obtained by
coupling the fields with a spacetime metric in such a manner that the
resulting curved spacetime action is invariant under local Weyl rescalings. The latter act on the metric as $g_{\mu\nu}\left(  x\right)  \rightarrow g_{\mu\nu
}^{\prime}\left(  x\right)  =\Omega^{2}\left(  x\right)  g_{\mu\nu}(x)$ and on
the matter fields as $\Phi_{I}\left(  x\right)  \rightarrow\Phi_{I}^{\prime
}\left(  x\right)  =\Omega^{s}\left(  x\right)  \Phi_{I}\left(  x\right)  $
where $s$ is the conformal weight of the matter field $\Phi_{I}$ and $I$
denotes some collective index structure. At the classical level, the field theories for a massless scalar, a massless Dirac fermion and a massless vector field with spin $s=0,\ 1/2,\ 1$, respectively, do possess conformal invariance. For a spin-2 field $h_{\mu\nu}\left(  x\right)$, a conformally invariant graviton action also exists and it is given by (see e.g. \cite{report} and references therein)
\begin{small}
\begin{align}\label{thelinaction}
S^{(2)}_{\mathrm{CG}}
=
\frac{1}{2}\int d^{4}x
\left[
(\Box h^{\mu\nu})(\Box h_{\mu\nu})
-2\,\Box h^{\mu\nu}\,\partial_{\mu}\partial^{\rho}h_{\rho\nu}
+\partial^{\mu}\partial^{\nu}h_{\mu\nu}\,
\partial^{\rho}\partial^{\lambda}h_{\rho\lambda}
-\frac{1}{3}
\left(
\Box h-\partial_{\mu}\partial_{\nu}h^{\mu\nu}
\right)^2
\right]
\end{align}
\end{small}
where \(h = \eta^{\mu\nu} h_{\mu\nu}\) and \(\Box = \partial^\mu \partial_\mu\),
leading to fourth-order field equations. The infinitesimal action of the global conformal group on $h_{\mu\nu}$ is given by
\begin{align}
\delta h_{\mu\nu}
&=\mathcal{L}_\epsilon h_{\mu\nu},
\end{align}
with
\begin{align}
\epsilon^\mu(x)
&= a^\mu
+ \omega^\mu{}_{\nu} x^\nu
+ \lambda x^\mu
+ \left( 2 (b \cdot x)\, x^\mu - b^\mu x^2 \right)\ .
\end{align}
Notice that the action \eqref{thelinaction} is also invariant under the gauge redundancies given by general diffeomorphisms $\delta_\text{diff}h_{\mu\nu}=\nabla_\mu\xi_\nu(x)+\nabla_\nu\xi_\mu(x)$ and local Weyl rescalings $\delta_\text{Weyl}h_{\mu\nu}=\sigma(x)\eta_{\mu\nu}$, both of which must be gauge-fixed when performing a functional integral.

The unique consistent completion of this theory to four-dimensional curved
spacetimes, invariant under local Weyl rescalings is \cite{Boulanger:2001he}
\begin{equation}\label{C2}
I\left[  g_{\mu\nu}\right]  =\alpha\int d^{4}x\sqrt{-g}C_{\ \ cd}^{a
b}C_{\ \ ab}^{cd}\ ,
\end{equation}
where $\alpha$ is a dimensionless coupling. This theory has been
considered as unphysical in its own merit, since it leads to fourth-order
field equations. Actually, the Euler-Lagrange equations of four-dimensional
conformal gravity are%
\begin{equation}
B_{\ b}^{a}\equiv\left(  \nabla^{c}\nabla^{d}+\frac{1}{2}R^{cd}\right)
C_{\ cbd}^{a}=0\ ,
\end{equation}
where $B_{\ b}^{a}$ is the Bach tensor. This tensor is also an obstruction
tensor for a spacetime to be conformally related to an Einstein
manifold, namely, a necessary condition for a spacetime to be conformal to an
Einstein manifold is the vanishing of the Bach tensor. This
condition is necessary but not sufficient since there are Bach-Flat
spacetimes which are not conformally Einstein like for example the
Nurowski-Plebanski spacetimes \cite{Nurowski:2000cq} (see also \cite{Liu:2013fna}). In spite of the
pathologies of its dynamical content, this theory has remarkable properties, as for
example it admits supersymmetric extensions (see e.g. \cite{report}), it emerges from
twistor-string theory \cite{Berkovits:2004jj} which implements superconformal
symmetry on the target space, it admits a Birkhoff's theorem \cite{Riegert:1984zz} (up to
conformal Weyl rescalings, of course) leading to black holes that have
topological extensions \cite{Klemm:1998kf}, and under suitable Neumann boundary conditions in
asymptotically AdS spacetimes it shares the same spectrum as General Relativity
\cite{Maldacena:2011mk} (see also \cite{Anastasiou:2016jix} and
 \cite{Grumiller:2013mxa} for holographic explorations in this theory). Even more, different attempts to construct a consistent quantum theory from
Conformal Gravity have been pushed forward \cite{Stelle:1976gc,Mannheim:2011ds,Oda:2023atd}. Conformal Gravity has also been useful to reobtain energy functionals in General Relativity as the Hawking mass and Willmore's energy functional in asymptotically AdS spaces \cite{Anastasiou:2022ljq}. A generic quadratic theory of gravity leading to fourth-order field equations
in dimension four, propagates a massless and a massive spin-2 mode, and a
scalar mode. Due to the invariance under conformal rescaling of the metrics,
the scalar mode of Weyl conformal gravity is non-dynamical, while the massive
mode becomes a ghost. When breaking the conformal invariance of the theory by
the introduction of the lower-derivative Einstein and cosmological
terms, the theory also possesses interesting properties. The theory admits
hairy black holes which must be constructed numerically
\cite{Lu:2015cqa,Lin:2016kip,Sajadi:2020axg,Sajadi:2025nkm} and can be also exactly determined from a closed form recurrence
relation \cite{Podolsky:2018pfe,Pravdova:2023nbo}. When formulated on AdS
spacetime, for a precise non-perturbative relation between the Weyl coupling
$\alpha$ and the cosmological constant, the linearized fourth-order operator
acting on $h_{\mu\nu}$ degenerates into the square of the single, well-defined,
second-order Lichnerowicz operator, leading on top of the GR modes also to log-modes in the so-called Critical Gravity \cite{Lu:2011zk} (see also \cite{Nutma:2012ss,Kleinschmidt:2013tsl} for the polycritical extensions). The
theory, at this special point and beyond can be naturally coupled to matter, conformal or not, leading to
interesting solutions (see e.g. \cite{Cisterna:2021xxq,Alvarez:2022wcj,Corral:2025npd}).

More recently, in \cite{BenettiGenolini:2026qdm}, the authors used conformal supergravity as a framework for higher-derivative supergravity theories, and used the off-shell formalism to study holographic observables in this scenario. It is worth mentioning that one can formulate conformal gravities in dimension larger than four by introducing non-analytic powers of a scalar field or raising the four-dimensional conformal gravity Lagrangian to the power $D/4$, again leading to a non-analytic Lagrangian for $D\neq 4n$ with $n\in\mathbb{Z}_{>0}$ (see e.g. the recent \cite{Hell:2026pwm}). Finally, the authors of \cite{Lescano:2023pai} show that the Double Copy provides a unified framework in which theories such as Conformal Gravity and Double Field Theory can be understood from gauge theoretic constructions.

As mentioned above, the field equations of four-dimensional conformal gravity vanish identically on Einstein manifolds, and therefore can complement the Einstein-Hilbert action without spoiling the physics of test fields on Schwarzschild spacetime. A unique, similar higher-derivative conformally invariant combination exists in six dimensions as shown by Lü, Pang and Pope in \cite{Lu:2013hx}, whose Noether-Wald charges were constructed in \cite{Anastasiou:2021tlv}. Recently, in \cite{Boulanger:2025oli} a similar unique construction has been given for arbitrary even dimensions, hinting the path to complete the program of Conformal Renormalization of holographic observables of codimension one and beyond pushed forward in \cite{Anastasiou:2026jrt}.

In this paper, we work within the realm of conformal invariant theories of gravity in even dimensions $D=2k$, constructed with a local Lagrangian, namely with invariants which are complete contractions of $k$ Weyl tensor in dimension $D=2k$. We study the thermodynamic properties of the unique, spherically symmetric black holes, as well as their topological extensions and show that {in every dimension there is a black-hole/black-hole zeroth-order phase transition at a given temperature}, which we consider as the main result of this work. The black holes are characterized by four constants which are restricted by a single algebraic constraint. {There is one integration constant controlling the leading asymptotic behavior of the metric, that becomes fixed if we fix the curvature radius of the asymptotic, locally AdS region. The subleading terms in the $r\to\infty$ expansion are universal, controlled by the remaining integration constants and are given by a linear term in the radial coordinate, then a subleading $\mathcal{O}(1)$ term and finally the first decaying subleading term at infinity which goes universally as $1/r$, regardless of the dimension}. We interpret this feature as the gravitational counterpart of the result in conformal electromagnetism in arbitrary dimensions, where the electric monopole potential decays as $1/r$ for every dimension as well \cite{Hassaine:2007py}. In spite of these slow fall-offs as compared with those of General Relativity \cite{Henneaux:1985tv}, we are able to obtain finite charges for every dimension, and extract physical consequencess from the thermodynamic quantities we obtain, which satisfy the first law of black hole thermodynamics.

The paper is organized as follows: In Section \ref{sec2}, we first use the covariant phase space method to obtain the mass of the four-dimensional black hole solution. We then verify the first law of thermodynamics and study the thermodynamic phase transitions. In Section \ref{sec3}, we extend our analysis to higher-dimensional black hole solutions with the same asymptotic behavior as the four-dimensional solution in conformal gravity. Finally, in Section \ref{seccon}, we present our conclusions and discuss possible directions for future work.

\section{Conformal Gravity in 4D}\label{sec2}

We begin with a review of the four-dimensional case, which has been addressed already in \cite{Lu:2012xu} and \cite{Peng:2014gha}\footnote{See also  \cite{Bekir} for a focus on the asymptotically de Sitter case, and the recent \cite{PenroseCG} for the study of the Penrose process on the rotating black hole of conformal gravity.}. As shown below, our analysis is complementary to the results of the latter references, with regard to the study of the four-dimensional thermodynamic quantities as well as of local and global thermal stability.

The action of conformal gravity in four dimensions is given in equation \eqref{C2}. For any diffeomorphism-invariant theory, depending on the curvature but not on its derivatives, variation of the action with respect to the metric leads to the following field equations
\begin{equation}
\mathcal{E}_{ab}\equiv\mathcal{P}_{a}{}^{cde}\,R_{bcde}
- 2\,\nabla^{c}\nabla^{d}\,\mathcal{P}_{acdb}
- \tfrac{1}{2}\, g_{ab}\, \mathcal{L}=0\ ,
\end{equation}
 where for the action \eqref{C2} one obtains
\begin{equation}
\mathcal{P}^{abcd}\equiv
\frac{\partial \mathcal{L}}{\partial R_{abcd}}
=2C^{abcd}.
\end{equation}
and we have set $\alpha=1$. The line element of a general spherically symmetric spacetime in four dimensions can be written as
\begin{equation}
    ds^2=g(t,r)\left[-f(t,r)dt^2+\frac{dr^2}{f(t,r)}+r^2d\Omega^2\right]
\end{equation}
where $d\Omega$ is the line element of the two-sphere. In the context of Conformal Gravity, the conformal factor $g(t,r)$ can be gauged-away. As shown by Riegert in \cite{Riegert:1984zz}, the field equations imply the existence of an additional Killing vector $\partial_t$, that is timelike when $f(t,r)=f(r)>0$, and therefore Birkhoff's theorem is fulfilled in Conformal Gravity, namely: spherical symmetry implies the existence of an extra Killing field which is timelike in the domain of outer communications of a black hole. This statement is also sometimes imprecisely summarized as \textit{``spherical symmetry implies staticity"} (see Appendix B of \cite{Hawking:1973uf} for a careful discussion of the theorem)\footnote{\linespread{0.85}\selectfont In this regard, Conformal Gravities belong to special family of theories fulfilling Birkhoff's theorem, the family being composed by General Relativity and Lovelock gravity \cite{Lovelock:1969vyr,Lovelock:1971yv}, Quasitopological gravities \cite{Oliva:2010zd,Myers:2010ru} and Conformal Gravities of the form $\text{Weyl}^n$ and their dimensional continuation  \cite{Oliva:2011xu}.}. The spacetime metric that solves the field equations for Conformal Gravity, in a suitable gauge takes the form
\begin{equation}
    ds^2=-f(r)dt^2+\dfrac{dr^2}{f(r)}+r^2d\theta^2+r^2\sin^{2}{\theta}d\phi^2
\end{equation}
where, the lapse function is
\begin{equation}\label{metric5}
f(r) = c + \frac{d}{r} + b\,r + a\,r^2.
\end{equation} provided the following constraint between the integration constants is satisfied\footnote{\linespread{0.85}\selectfont This solution was first obtained by Riegert \cite{Riegert:1984zz} and subsequently studied by Mannheim and Kazanas \cite{Mannheim:1988dj} in the context of galactic velocity curves.}
\begin{equation}\label{eqcond}
    3bd=c^2-1.
\end{equation}
The asymptotic behavior of the spacetime is controlled by the constant $a$. For asymptotically AdS spacetimes $a=l^{-2}$, where $l$ is the curvature radius of the asymptotic region\footnote{\linespread{0.85}\selectfont Notice that by introducing a suitable, improper Weyl rescaling (conformal gluing), one can construct wormhole geometries \cite{wormhole4d}, which can also support the field of an electric and a magnetic monopole, without source \cite{Hohmann:2018shl}.}. The linear term in $r$ is a feature specific to Conformal Gravity and we would like to study the thermal properties of the black holes included in this family of geometries when this proper feature is present, namely when $b\neq0$, unless otherwise stated. Consequently, we solve the restriction \eqref{eqcond} for the integration constant $d$ such that $d=\frac{c^2-1}{3b}$.

For diffeomorphism-invariant theories with a Lagrangian depending on the Riemann tensor, \(\mathcal{L}(g,\mathcal{R})\), the symplectic potential takes the form
\begin{equation}
\label{wald-potential-formula}
\Theta^{a}
=
2 \mathcal{P}^{abcd}\,\nabla_{c}h_{bd}
-
2(\nabla_{c}\mathcal{P}^{abcd})\,h_{bd},
\end{equation}
here $h_{ab}=\delta g_{ab}$, namely an infinitesimal variation in the functional space of metrics. For the Riegert solution the only non-vanishing component of the symplectic potential is
\begin{equation}
\label{wald-symplectic}
 \Theta^{r}= -\dfrac{4(c-1)}{3}\left(\dfrac{c+1}{b}\,\,\delta a-\,\delta b\right)\sqrt{\omega}\ ,  
\end{equation}
where $\omega$ is the determinant of the metric of the two-sphere $S^2$. The Noether current associated with a vector field $\xi^{a}$ can be written as
$J^{a} = \nabla_{b} Q^{ab}$,
where the antisymmetric Noether charge is
\begin{equation}
\label{wald-noethercurrent}
Q^{ab}
=
-2\, \mathcal{P}^{abcd}\, \nabla_{c}\xi_{d}
+ 4\, \xi_{d}\, \nabla_{c} \mathcal{P}^{abcd}\ ,
\end{equation}
which for Riegert solution possesses a single non-vanishing component, given by
\begin{equation}
    Q^{tr}=-\dfrac{4(c-1)}{3}\left(\dfrac{2a(c+1)}{b}-b(c-1)\right)\sqrt{\omega}.
\end{equation}
In consequence the variation of the Hamiltonian
associated with $\xi^{a}$ is given by
\begin{equation}
\delta H_\xi=\int_{\partial\Sigma}k^{a b}\,dS_{ab}
=
\int_{\partial\Sigma}
\Big(
\delta Q^{ab}_\xi
-
\xi^{[a}\Theta^{\,b]}(h)
\Big)
\, dS_{ab}.
\end{equation}

Evaluating for $\xi=\partial_t$ and integrating over the sphere at infinity under the condition \eqref{eqcond} leads to
\begin{align}
\delta M = \oint_{S^{2}_\infty} k^{tr}\, dS_{tr}=&\frac{16\pi}{3}
\frac{1}{b^2}\Big[(b^3-4\,c\,a\,b)\,\delta c+2\,(c^2-1)\,a\,\delta b-(c^2-1)\,b\delta a\Big].
\end{align}
As can be seen, this expression is not integrable in general. For our purposes, we focus on the case in which integrability is achieved by keeping both the curvature scale of the asymptotic region $l=1/\sqrt{a}$ ($a>0$) and the subleading term controlled by the constant $b$, which is linear in $r$, fixed. Therefore, hereafter we impose $\delta a=\delta b=0$. Since we are considering $b\neq 0$, the solutions under consideration are asymptotically locally AdS, approaching AdS in a relaxed manner with respect to the Henneaux-Teitelboim \cite{Henneaux:1985tv} and Henneaux \cite{HenneauxHigherD} conditions. Consequently, the AdS spacetime itself cannot be used as the reference configuration within the family considered here. For the class of solutions with $b\neq 0$, we choose as the background reference the spacetime where the symplectic potential \eqref{wald-symplectic} and the Noether potential \eqref{wald-noethercurrent} identically vanish, which corresponds to impose $c=1$. In that case, it can be shown that the resulting spacetime is conformally related to AdS$_2\times \mathbb{S}^2$. In sum, the mass of the black hole is given by
\begin{align}
    \label{mass-dim4}
M&= \dfrac{16\pi (c-1)}{3b}(b^2-2ac-2a)\,.
\end{align}
This quantity is in agreement with the conserved charge associated with the timelike Killing vector, constructed from the holographic response functions of Conformal Gravity \cite{Grumiller:2013mxa}. The choice $c=1$ as the reference state can also be implemented directly within the Quasilocal Formalism. In this approach, the ADT potential $Q_{ADT}^{\mu\nu}$ is related to the Noether potential $Q^{\mu\nu}$ and the symplectic potential $\Theta^\mu$ according to \cite{Kim:2013zha,Gim:2014nba}
\begin{equation}
\sqrt{-g}Q_{ADT}^{\mu\nu}
=\frac{1}{2}\delta\left(\sqrt{-g}Q^{\mu\nu}\right)
-\sqrt{-g}\xi^{[\mu}\Theta^{\nu]},
\end{equation}
where $\xi$ is a Killing vector. Introducing a one-parameter path in the space of solutions, $g_{\mu\nu}(s)$, with $s\in[s_0,s_1]$, the quasilocal charge associated with $\xi$ can be written as
\begin{equation}
\mathcal{Q}(\xi)
=\int d^{D-2}x_{\mu\nu}
\left(
\Delta Q^{\mu\nu}(\xi)
-2\xi^{[\mu}\int_{s_0}^{s_1}ds\Theta^{\nu]}(s)
\right).
\end{equation}
Choosing the initial point of the path, $g_{\mu\nu}(s_0)$, to be the solution with $c=1$, one checks that the quasilocal mass coincides with $M$ from Eq.~\eqref{mass-dim4}.

\subsection{Thermodynamics and Stability of the Riegert Black Hole}
\begin{figure}
\centering
\includegraphics[width=0.45\columnwidth]{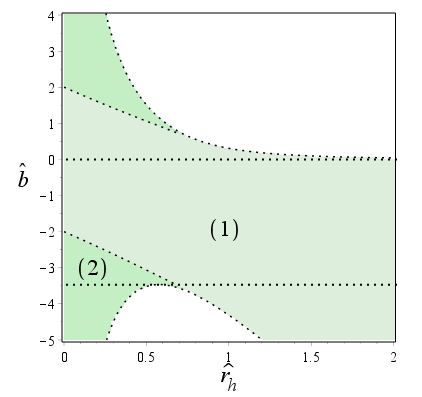}
\includegraphics[width=0.45\columnwidth]{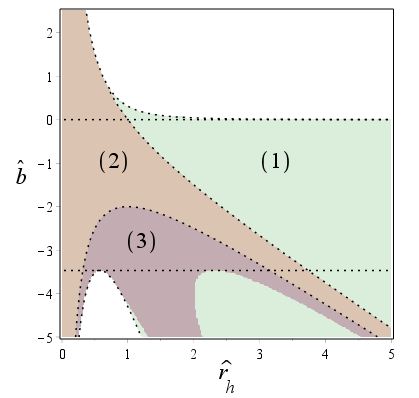}
\caption{\textbf{Phase diagram in the $(\hat{r}_h,\hat{b})$ parameter space}. Left panel: The shaded regions labeled $(1)$ and $(2)$ correspond to black hole configurations with one horizon and three real positive roots of $f(r)$, respectively, for the branch $c^{+}$. Right panel: The shaded regions labeled $(1)$, $(2)$, and $(3)$ correspond to black hole configurations with one horizon, two real positive horizons, and three real positive roots of $f(r)$, respectively, for the branch $c^{-}$. The white regions are excluded from the physical parameter space, as they correspond to negative or complex values of the relevant parameters. Notice that here $r_h$ represents the largest of the possible up to three zeros of the function $f(r)$.}
\label{fig:metric}
\end{figure}

{We now investigate the thermodynamic properties of the black holes described in the previous section}. In particular, we derive the Hawking temperature, entropy, mass, and examine the validity of the first law of black hole thermodynamics within the framework of Weyl conformal gravity. The equation $f(r)=0$ with $f(r)$ given in \eqref{metric5} may possess more than one solution. Since we are restricting to asymptotically AdS spacetimes, the largest root $r=r_h$ of $f(r)$ defines an event horizon. The Hawking temperature associated with the event horizon, requiring regularity of the Euclidean continuation, as usual is given by
\begin{equation}
T=\frac{f^{\prime}(r_{h})}{4\pi}
=\frac{6abr_{h}^3+3b^2r_{h}^2-c^2+1}{12\pi b r_{h}^2}.
\end{equation}
Imposing the horizon condition, $f(r_{h})=0$, one finds that the integration constant $c$ can be expressed as
\begin{equation}\label{eqehc}
c^{\pm}
=-\frac{3br_{h}}{2}
\pm \frac{\sqrt{\Delta}}{2},
\qquad
\Delta
=4-12abr_{h}^{3}-3b^{2}r_{h}^{2}.
\end{equation}
Figure~\ref{fig:metric} illustrates the phase structure of the black hole solutions in the $\left(\hat{r}_h=\sqrt{a}\,r_{h}=\frac{r_h}{l}, \hat{b}=\frac{b}{\sqrt{a}}=l b\right)$ parameter space. For the $c^{+}$ branch, the physical parameter space is divided into two regions corresponding to black holes with one and three horizons located to the right of the curvature singularity at $r=0$. In contrast, the $c^{-}$ branch exhibits a richer structure, containing regions with one, two, and three horizons.

The requirement of $c$ being real imposes the condition $\Delta \geq 0$, which constrains the allowed parameter space of the black hole solutions.
Substituting the positive and negative branches, $c=c^{\pm}$, into the expression for the temperature yields
\begin{equation}
T=
\frac{
b r_{h}
+6a r_{h}^{2}
\pm\sqrt{\Delta}
}{8\pi r_{h}}\ ,
\end{equation}
with $\Delta$ defined in \eqref{eqehc}.

For the Riegert black hole, the Wald entropy is given by
\begin{equation}
S_{W}
=-2\pi \int_{H} d^{2}x\sqrt{h}
P^{abcd}\epsilon_{ab}\epsilon_{cd}
=\frac{64\pi^2 \alpha}{3}
\left[
-\frac{r_h^2}{2}f''(r_h)
+r_h f'(r_h)
+1
\right],
\end{equation}
where $H$ denotes the bifurcation surface of the event horizon, $h$ is the determinant of the induced metric on $H$, and $\epsilon_{ab}$ is the binormal to the horizon surface. Substituting the metric function \eqref{metric5} into the above expression and using $c=c^{\pm}$ as given in \eqref{eqehc}, we obtain
\begin{equation}
S =
\frac{64\pi^2\alpha}{3\,b\,r_{h}}
\left[
(1-c)(c+1+b\,r_{h})\right]=
\frac{16\pi^{2}\alpha}{3 b r_{h}}
\left(2+3 b r_{h}\mp\sqrt{\Delta}\right)
\left(2-b r_{h}\pm\sqrt{\Delta}\right).
\end{equation}
Using the thermodynamic quantities obtained above, one can verify that the first law of thermodynamics is satisfied in the form \cite{Peng:2014gha}
\begin{equation}
dM=TdS\,.
\end{equation}
It should be noted that here we hold $a$ and $b$ fixed, and that infinitesimal variations within equilibrium configurations implied in the first law, stand for variations with respect to $r_h$, only.\\

\begin{figure}
\centering
\includegraphics[width=0.45\columnwidth]{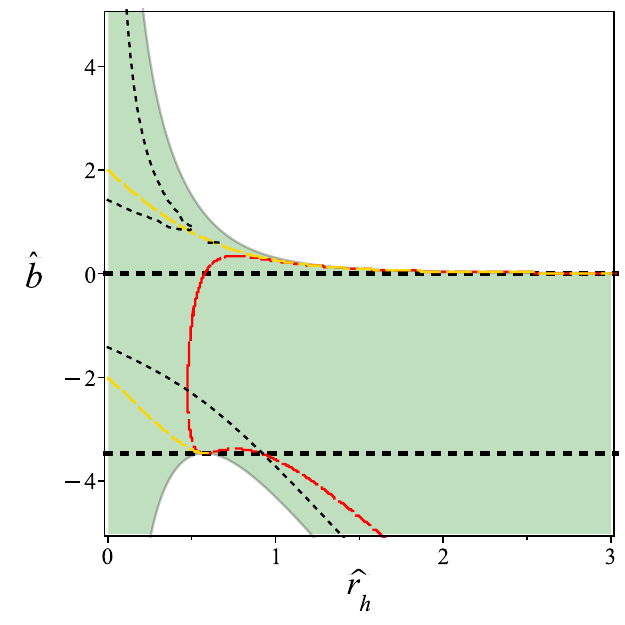}
\includegraphics[width=0.45\columnwidth]{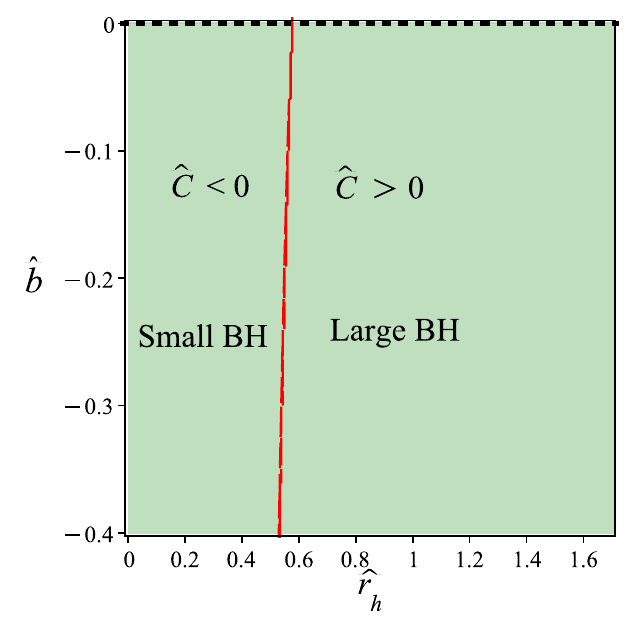}
\caption{\textbf{Phase diagram in the $(\hat{r}_h,\hat{b})$ parameter space for the case of $c^{+}$.} The green-shaded region corresponds to values of the parameters for which $\hat{\Delta}$ is real, while the white region is excluded from the physical parameter space because it yields complex-valued quantities. The region in which the black hole can be extended to infinity is bounded by the two dotted curves. The red curve is determined by the condition $\hat{T}^{\prime}=0$ and separates the small- and large-black-hole branches. The region enclosed by the dashed yellow curves and the red curve is characterized by negative heat capacity, whereas the remaining physically allowed regions have positive heat capacity (left). The right panel shows an enlarged view of a portion of the parameter space (right). 
}
\label{fig:phase}
\end{figure}

We next study the thermodynamic stability and possible phase transitions for asymptotically AdS black holes, in the canonical ensemble. Namely, for fixed $a$ and $b$, we determine which phase minimizes the Helmholtz Free Energy $F(T)=M-TS$, while keeping track of the sign of the heat capacity
\begin{equation}
C=T\frac{dS}{dT}\ ,
\end{equation}
for each black hole branch. Notice also that, since we are interested in keeping a fixed non-vanishing parameter $b$ in \eqref{metric5}, it will not be possible to reach the thermal AdS solution of the theory which has $b=0$ and curvature radius $l=a^{-1/2}$. Therefore, we will evaluate differences of free-energies between different thermal black hole configurations to determine whether it would be possible to have a change in the dominating black hole phase, namely a black hole-black hole phase transition. To this end, it is convenient to introduce the following dimensionless variables which allow us to explore in a simpler manner the physical regions in the parameter space:
\begin{equation}\label{eqqdim}
    \hat{b}=lb, \qquad
    \hat{r}_{h}=\frac{r_{h}}{l},
    \qquad a>0\ .
\end{equation}
Accordingly, $\hat{\Delta}$ can be rewritten as follows
\begin{align}\label{cond1}
\hat{\Delta}=-3\hat{b}^2\hat{r}_h^2-12\hat{b} \hat{r}_h^3+4\ge0,\quad\quad\quad \hat{r}_h>0\ .
\end{align}
The admissible region in the $(\hat{r}_h,\hat{b})$ parameter space is determined by the following inequalities:
\begin{equation}
-2\hat{r}_{h}-\frac{2}{\sqrt{3}}
\sqrt{3\hat{r}_{h}^{2}+\frac{1}{\hat{r}_{h}^{2}}}
\leq \hat{b}\leq
-2\hat{r}_{h}+\frac{2}{\sqrt{3}}
\sqrt{3\hat{r}_{h}^{2}+\frac{1}{\hat{r}_{h}^{2}}},\qquad \hat{r}_h>0
\end{equation}
However, black hole solutions with horizon radii extending over the full range $0\leq \hat{r}_{h}<\infty$
exist only when the parameter $\hat{b}$ satisfies $-2\sqrt{3}\leq \hat{b}\leq 0.$
Figure~\ref{fig:phase} illustrates the resulting phase structure in the $(\hat{r}_h,\hat{b})$ plane. The green region corresponds to real $\hat{\Delta}$.

Using these dimensionless variables, the thermodynamic quantities can be rewritten in the following dimensionless form:
\begin{align}
    \hat{T}^{\pm}=\dfrac{T^{\pm}}{\sqrt{a}}
    =&\frac{\hat{b}\hat{r}_{h}+6\hat{r}_{h}^{2}
    \pm\sqrt{\hat{\Delta}}}
    {8\pi \hat{r}_{h}},\\
\hat{M}^{\pm}=\sqrt{a}M^{\pm}=&\frac{8\pi}{3\hat{b}}
\left(
-3\hat{b}\hat{r}_{h}
\pm\sqrt{\hat{\Delta}}
-2\right)
\left(
\hat{b}^{2}+3\hat{b}\hat{r}_{h}
\mp\sqrt{\hat{\Delta}}
-2\right),\\
\hat{S}^{\pm}
=&\frac{16\pi^{2}}{3\hat{b}\hat{r}_{h}}
\left(-
3\hat{b}\hat{r}_{h}
\pm\sqrt{\hat{\Delta}}
-2
\right)
\left(
\hat{b}\hat{r}_{h}
\mp\sqrt{\hat{\Delta}}
-2
\right).
\end{align}

The temperature is real and positive provided the parameter $\hat{b}$ satisfies $-3\hat{r}_{h}-\frac{1}{\hat{r}_{h}} \leq \hat{b} \leq 0,\, \hat{r}_{h}\geq 0.$ To determine the minimum temperature, we impose
\begin{equation}
\frac{d\hat{T}}{d\hat{r}_{h}}=0.
\end{equation}
This condition yields
\begin{equation}
    \hat{b}_{\text{min}}\hat{r}_h^3 - \hat{r}_h^2\sqrt{-3\hat{b}_{\text{min}}^2\hat{r}_h^2-12\hat{b}_{\text{min}} \hat{r}_h^3+4} + \dfrac{2}{3}=0\,,\quad\quad\text{and}\quad\quad \hat{r}_{h}\geq 0.47.
\end{equation}

The behavior of $\hat{b}_{\min}$ in terms of $\hat{r}_{h}$ is shown in Fig.~\ref{fig:phase} by the red dashed curve. This curve separates the parameter space into regions corresponding to small and large black holes. The region to the left of the red curve, associated with small black holes, has negative heat capacity and is therefore locally thermodynamically unstable, whereas the region to the right, corresponding to large black holes, has positive heat capacity and they are thermodynamically stable.
Defining
\begin{equation}
\hat{\mathbf{A}}
=\sqrt{81\hat{r}_{h}^{8}+54\hat{r}_{h}^{4}-3}\ ,
\end{equation}
the corresponding minimum temperature can be written as
\begin{equation}
\hat{T}_{\min}=
\frac{1}{48\pi \hat{r}_{h}^{3}}
\left[
27\hat{r}_{h}^{4}
+\hat{\mathbf{A}}
-1
+\sqrt{6\left(
27\hat{r}_{h}^{8}
-3\hat{r}_{h}^{4}\hat{\mathbf{A}}
+\hat{\mathbf{A}}
+1
\right)}
\right].
\end{equation}
Therefore, black hole solutions exist only for temperatures satisfying
\begin{equation}
\hat{T}\geq \hat{T}_{\min},
\end{equation}
while no black-hole configuration exists below the minimum temperature, similar to what occurs for the Schwarzschild AdS black hole in General Relativity. Notice, nevertheless, that $T_\text{min}$ can vanish, and in such case the black holes are continuously connected with an extremal black hole.

We have obtained the mass, temperature, and entropy of the black holes
in conformal gravity. With these expressions, we can compute the free
energy. There are two branches of solutions depending on the value that the
integration constant $c$ takes in terms of the boundary condition $b$, and the
event horizon radius. For these two branches, one must compare their
free energy and select as the dominating phase in the canonical ensemble the
one with the lowest value of $F=M-TS$, at a fixed temperature.

\begin{figure}
\centering
\includegraphics[width=0.75\columnwidth]{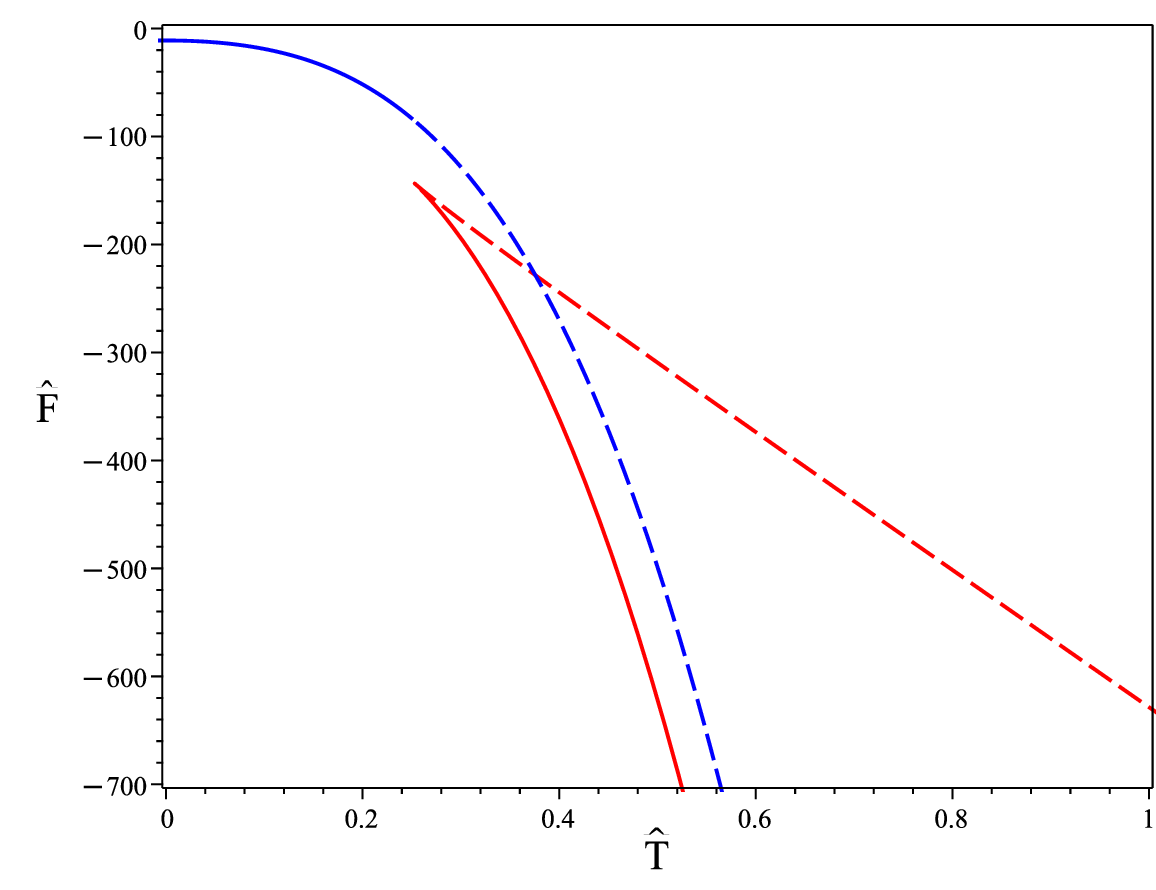}
\caption{Free energy $\hat{F}$ as a function of temperature $\hat{T}$ for two distinct black-hole solution branches (blue and red), with the same value for the parameter $b=-1$.}
\label{fig:ftplot}
\end{figure}

It is useful to fix the value of $b$ (fixed by boundary
conditions) and parametrize the curve $\left(  F,T\right)  $ as a function of
parameter $r_{h}$. We have seen that for different ranges of the temperature
and values of $b$, there can be from none to up to three black holes. Notice that
since we are working with a fixed, non-vanishing value of $b$, the metric
function always has a subleading term $r^{2}/l^2+br+...$, which cannot be turned
off, therefore thermal AdS is not an allowed phase for non-vanishing $b$. Below we present the free energy versus
temperature plots for $\hat{b}=-1$ in Figure \ref{fig:ftplot}.  At low temperatures, the blue branch (small black hole) represents the thermodynamically stable phase. As the temperature increases, the red branch becomes the favored branch, exhibiting a lower free energy than the blue branch and thus taking over as the stable phase. The two branches are separated by a finite gap in free energy, and no smooth crossing point exists between them, indicating that the transition between the phases occurs via a discontinuous change rather than through a continuous equality of free energies. This behavior is characteristic of a zeroth-order phase transition between black-hole phases. Transitions of this kind have already been observed in black hole physics, as for example in \cite{Cong:2021jgb}.

\section{Thermodynamics of black hole in higher-dimensional conformal gravity}\label{sec3}

Consider the action in $D = 2k$ dimensions:
\begin{equation}\label{Ik}
I^{(k)} = \int d^{2k}x \sqrt{-g}\, \alpha^{(k)} W^{(k)},
\end{equation}
where there is no sum in $k$ and we have defined
\begin{equation}
W^{(k)} = C_{ab}{}^{cd} C_{cd}{}^{ef} \cdots C_{mn}{}^{ab}
\quad (k\ \text{Weyl tensors}).
\end{equation}
Notice that this expression can be rewritten as a trace of $k$ Weyl tensor, namely $W^{(k)}=tr(C^k)$. In order to contruct conformal gravitites that are algebraic in the Weyl tensor, one can consider different, algebraically independent contractions of $k$ Weyl tensors in dimension $D=2k$, or even a generic linear combination of them. Actually, in $D=6$, there are two independent algebraic conformal gravities, while in $D=8$, there are seven different independent choices (see e.g. the seminal work \cite{Fulling:1992vm}). For spherically symmetric spacetimes, the field equations of all these conformal gravities reduce to the same expression, and therefore, we restrict the analysis to a single representative. At the spherically symmetric level, considering different conformally invariant theories amounts to a redefinition of the coupling $\alpha^{(k)}$, provided one is not considering the critical case at which the linear combination of different conformal gravities may cancel each other. Notice that once one moves beyond spherical symmetry, this equivalence will no longer hold, and different conformal gravities may lead to different physics.  

Again, variation of the action \eqref{Ik} with respect to the metric gives the following result:
\begin{equation}
\mathcal{E}_{ab}=\mathcal{P}_{a}{}^{cde}\,R_{bcde}
- 2\,\nabla^{c}\nabla^{d}\,\mathcal{P}_{acdb}
- \tfrac{1}{2}\, g_{ab}\, \mathcal{L}=0 \ .
\end{equation}
By expressing the Weyl tensor in terms of the Riemann tensor, Ricci tensor, and Ricci scalar, and subsequently employing the symmetries and identities of the Weyl tensor, one arrives at the following general expression for the Wald tensor in conformal gravity in dimension $D=2k$:  
\begin{align}\label{eqqwaldtensor}
P^{(k)}_{abcd}\equiv
\frac{\partial \mathcal{L}^{(k)}}{\partial R_{abcd}}
&=
k\,\alpha^{(k)}\Big[(C^{(k-1)})_{abcd}
-\frac{2}{D-2}\left(g_{a[c}A^{(k-1)}_{d]b}-g_{b[c}A^{(k-1)}_{d]a}
\right)\nonumber \\
&+\frac{2}{(D-1)(D-2)}
B^{(k-1)} g_{a[c}g_{d]b}
\Big],
\end{align}
where
\begin{equation}
\label{AB}
A^{(k-1)}_{ab} = (C^{(k-1)})_{acb}{}^{c},\quad B^{(k-1)} = (C^{(k-1)})_{ab}{}^{ab}.
\end{equation}
For $k=2$, using the tracelessness of the Weyl tensor, one finds that $A=0$ and $B=0$. Therefore, only the first term in the above expression survives, reproducing the result obtained in the previous section.

Birkhoff's theorem proved by  Riegert in four-dimensional conformal gravity, was extended to arbitrary even dimensions in \cite{OlivaRay2012}. The general black hole solution of the theory in this case is given by
\begin{equation}
ds^2 = 
-f(r) dt^2 + \frac{dr^2}{f(r)}
+ r^2 d\Sigma_{\gamma, D-2}^2,
\end{equation}
with
\begin{equation}\label{eqhid}
f_\text{original}(r) = \tilde{a} \,r^2 + 2 \,\tilde{b}\gamma \,r + \gamma
+ \tilde{c} \,r^2\left(\frac{1}{r} + \tilde{b}\right)^{\frac{2k-1}{k-1}}\,,
\end{equation}
as originally presented in \cite{OlivaRay2012}. Here, $d\Sigma_{\gamma, D-2}^2$ represents the metric on the $(D-2)$-dimensional space of constant curvature $\gamma$, normalized to $1$ for a sphere, $0$ for a torus, or $-1$ for a hyperbolic space. We find it useful to introduce a reparametrization of the metric function $f(r)$, which hereafter we will consider to be given by

\begin{equation}\label{newf}
\begin{aligned}
f_\text{new}(r)=&
\left(
1-\frac{b^{2}(c-\gamma)(k-1)^{2}k}
{2(2k-1)\bigl(c(k-1)+\gamma\bigr)^{2}}
\right)r^{2}
+\frac{bk\gamma}{c(k-1)+\gamma}\,r+\gamma
\\
&
+\frac{(c-\gamma)(k-1)^{2}}
{2(2k-1)k}
\left(
\frac{bk}{c(k-1)+\gamma}
\right)^{-\frac{1}{k-1}}
r^{-\frac{1}{k-1}}
\left(
\frac{bkr}{c(k-1)+\gamma}+2
\right)^{\frac{2k-1}{k-1}} .
\end{aligned}
\end{equation}
Although this new parametrization seems more intricate, it is adapted to clearly unveil the asymptotic meaning of the integration constants. Indeed, for arbitrary dimension $D=2k$ we have
\begin{equation}\label{universalasymp}
f_\text{new}(r)=r^2+br+c+\frac{2(c-\gamma)(c(k-1)+\gamma)}{3bk(k-1)}\frac{1}{r}+\mathcal{O}(r^{-2})\ ,
\end{equation}
as $r\to\infty$. We have fixed the cosmological radius at infinity to 1, and have defined the integration constants in such a manner that we have independent control on the subleading term, linear in $r$, via the integration constant $b$. The power series truncates only in dimension $D=4$ for $k=2$, since $\frac{2k-1}{k-1}$ is an integer only in such a case. Although more intricate, this parametrization allows to study the phase space for different values of $c$, for a fixed dimension, and fixed curvature of the asymptotic radius and subleading term $b$. Notice that the asymptotic behavior always contains a subleading term that decays as the four-dimensional Newtonian potential. This feature can be interpreted as the gravitational analogue of what occurs in conformal electrodynamics, where the electric monopole behaves as $A_t\sim r^{-1}$, regardless of the dimension \cite{Hassaine:2007py}. Finally, notice that for $k>1$, a new type of singularity may emerge, namely a branch singularity, located at
\begin{equation}\label{branchpoint}
    r_\text{bs}=-\dfrac{2(c(k-1)+\gamma)}{bk}.
\end{equation}

For odd values of \(k\), \(k-1\) is even, and hence the exponent \((2k-1)/(k-1)\) in \eqref{newf} has an even denominator. Using \eqref{branchpoint} the corresponding factor in \eqref{newf} can be written as
\begin{equation}
\left[\frac{bk}{c(k-1)+\gamma}(r-r_{\mathrm{bs}})\right]^{\frac{2k-1}{k-1}}.
\end{equation}

Therefore, provided that \(bk/[c(k-1)+\gamma]>0\), reality of the metric requires \(r-r_{\mathrm{bs}}\geq0\). If \(r_{\mathrm{bs}}>0\), the spacetime consequently has a finite radial endpoint at \(r=r_{\mathrm{bs}}\), and the physical radial domain is
$r \in [r_{\mathrm{bs}},\infty)$.

This behavior is similar to the branch singularity that arises in charged Einstein--Gauss--Bonnet black hole solutions \cite{Torii:2005nh,Wiltshire:1985us}. The spacetime also has a curvature singularity at $r=0$ and we will deal with the cases in which both, the latter singularity and the potential branch singularity are covered by an event horizon, in an asymptotically locally AdS spacetime. The branch singularity, as soon as $r_\text{bs}>0$ will inevitably appear in a curvature invariant constructed with a sufficiently large number of derivatives of the Riemann tensor.

With the above discussion settled, we now turn into the computation of the Wald entropy for the higher-dimensional black hole solution. To that end, we choose to work in the following conformal frame,
\begin{equation}
    \label{metric-directsum}
    d\tilde{s}^2 \equiv -\dfrac{f(r)}{r^2} dt^2 + \frac{dr^2}{r^2f(r)} + d\Sigma_{\gamma, D-2}^2 = \dfrac{1}{r^2} ds^2\,,
\end{equation}
because, as shown in \cite{OlivaRay2012}, the Weyl tensor is determined by a single scalar $\tilde{S} = \tilde{R} + 2\gamma$, where $\tilde{R}$ denotes the scalar curvature of the two-dimensional space orthogonal to $\Sigma_{\gamma, D-2}$. Indeed,
\begin{equation}
\label{weyl-relations}
C_{jl}^{\ ik}=\dfrac{(D-3)\tilde{S}}{2(D-1)}\delta_{jl}^{ik}\,,\quad C_{\nu\rho}^{\ \mu\lambda}=\dfrac{\tilde{S}}{(D-1)(D-2)}\delta_{\nu\rho}^{\mu\lambda}\,,\quad 
C_{j\nu}^{\ i\mu}=-\dfrac{(D-3)\tilde{S}}{2(D-1)(D-2)}\delta_{j}^{i}\delta_{\nu}^{\mu}\,.
\end{equation}

To obtain the general formula for the entropy, recall that the binormal $\epsilon_{ab}$ lives in the $t$--$r$ plane, and it is normalized as $\epsilon_{ab}\epsilon^{ab}=-2$. Thus, the contraction $P^{abcd}\epsilon_{ab}\epsilon_{cd}$ collapses to one component only, $P^{abcd}\epsilon_{ab}\epsilon_{cd} = -4P_{tr}^{\ tr}$. The Wald entropy therefore reads
\begin{align}
S_{k} = -2\pi \int_{\mathcal{H}}d^{2k-2}x\sqrt{h}P^{abcd}\epsilon_{ab}\epsilon_{cd} = 8\pi \textrm{Vol}(\Sigma_{\gamma,2k-2}) P_{tr}^{\ tr}\Big|_{r=r_h}\,,
\end{align}
and since all the components of the Weyl tensor depend on $\tilde{S}$, $P_{tr}^{\ tr}$ is proportional to $\tilde{S}^{k-1}$, namely $P_{tr}^{\ tr} = k\mu_{2k}\tilde{S}^{k-1}$. It only remains to compute the quantity $\mu_{2k}$.

Working in the conformal presentation of the metric \eqref{metric-directsum}, the Weyl tensor is completely determined by \eqref{weyl-relations}. Equivalently, thought as a bivector transformation, one can say that the Weyl operator is diagonal on the space of two-forms, and its eigenvalues $\lambda_{[AB]}$ are read
directly off \eqref{weyl-relations}:
\begin{equation}
  \lambda_{[tr]}=\frac{2k-3}{2(2k-1)}\,\tilde S\,,\qquad
  \lambda_{[ab]}=\frac{1}{(2k-1)(2k-2)}\,\tilde S\,,\qquad
  \lambda_{[ta]}=-\frac{2k-3}{2(2k-1)(2k-2)}\,\tilde S\,,
  \label{eq:eigs}
\end{equation}
with multiplicities $1$, $\binom{2k-2}{2}$ and $2(2k-2)$ respectively. Seen as a diagonal operator, the $C^{(k-1)}$ chain acts on each eigen-bivector by a power,
\begin{equation}
  (C^{(k-1)})_{[AB]}{}^{[AB]}=2^{k-2}\,\lambda_{[AB]}^{\,k-1}
  \qquad(\text{no sum})\,,
  \label{eq:chain}
\end{equation}
and the factor $2^{k-2}$ counts the $k-2$ internal ordered contractions.

Recalling Eqs. \eqref{eqqwaldtensor} and \eqref{AB} in terms of $k$,
\begin{align}
P^{(k)}_{abcd}\equiv
\frac{\partial \mathcal{L}^{(k)}}{\partial R_{abcd}}
&=
k\,\alpha^{(k)}\Big[(C^{(k-1)})_{abcd}
-\frac{2}{2k-2}\left(g_{a[c}A^{(k-1)}_{d]b}-g_{b[c}A^{(k-1)}_{d]a}
\right)\nonumber \\
&+\frac{2}{(2k-1)(2k-2)}
B^{(k-1)} g_{a[c}g_{d]b}
\Big],
\end{align}
where
\begin{equation}
A^{(k-1)}_{ab} = (C^{(k-1)})_{acb}{}^{c},\quad B^{(k-1)} = (C^{(k-1)})_{ab}{}^{ab}\,,
\end{equation}
one finds
\begin{align}
  (C^{(k-1)})_{tr}^{\ \ tr}
  &=2^{k-2}\,\lambda_{[tr]}^{\,k-1}\,,
   \label{eq:Ctrtr}\\[4pt]
  (A^{(k-1)})_{t}^{\ t}+(A^{(k-1)})_{r}^{\ r}
  &=2^{k-1}\Big[\lambda_{[tr]}^{\,k-1}+(2k-2)\,\lambda_{[ta]}^{\,k-1}\Big]\,,
   \label{eq:AA}\\[4pt]
  B^{(k-1)}
  &=2^{k-1}\Big[\lambda_{[tr]}^{\,k-1}+(4k-4)\,\lambda_{[ta]}^{\,k-1}
        +(k-1)(2k-3)\,\lambda_{[ab]}^{\,k-1}\Big]\,.
   \label{eq:BB}
\end{align}
Inserting \eqref{eq:eigs} into \eqref{eq:Ctrtr}--\eqref{eq:BB} to compute $P_{tr}^{\ tr}$, one obtains
\begin{align}
P_{tr}^{\ tr} = k\mu_{2k}\tilde{S}^{k-1}\,,
\end{align}
where
\begin{equation}
\label{eq:mu}
  \mu_{2k}=\frac{1}{2\,(2k-1)^{k}}
  \left[(2k-3)^{k}
        +\frac{2k-3}{(k-1)^{k-1}}
         \left((-1)^k\frac{(2k-3)^{k-1}}{2^{k-2}}+1\right)\right].
\end{equation}

In terms of $\mu_{2k}$ the Wald entropy reads
\begin{align}
  S_{k}&=8\pi k\mu_{2k}\tilde{S}^{k-1}(r_{h})\,\textrm{Vol}(\Sigma_{\gamma, 2k-2})\,,
\end{align}
and using the explicit expression for $\tilde{S}(r_h)$, one finds
\begin{align}
\label{wald-entropy-k}
S_{k}
        &= \dfrac{2^{k+2}\pi \mu_{2k}}{br_h} (\gamma-c)^{k-1} (2\gamma +(2k-2)c+kbr_h) \textrm{Vol}(\Sigma_{\gamma, 2k-2})\,.
\end{align}

The first few members of the sequence $\mu_{2k}$ are
\begin{equation}
\mu_{4}=\frac{1}{6},\qquad
\mu_{6}=\frac{39}{400},\qquad
\mu_{8}=\frac{3245}{24696}\,,
\end{equation}
so the Wald entropy for the first few dimensions reads
\begin{align}
S_{2} &= \frac{16\pi}{3br_h}(\gamma-c)\left(\gamma+c+br_h\right)\,\textrm{Vol}(\Sigma_{\gamma, 2}),\\[1ex]
S_{3} &= \frac{78\pi}{25br_h}(\gamma-c)^2
\left(2\gamma+4c+3br_h\right)\,\textrm{Vol}(\Sigma_{\gamma, 4})\\[1ex]
S_{4} &= \frac{51920\pi}{3087br_h}(\gamma-c)^3
\left(\gamma+3c+2br_h\right)\,\textrm{Vol}(\Sigma_{\gamma, 6})\,.
\end{align}

Proceeding as before, one can compute the mass of these black holes by fixing the value of the subleading term $\delta b=0$. In the new parametrization $f_{new}(r)$ \eqref{newf}, the spacetime with $c=\gamma$ is conformal to a direct product between AdS$_2$, $\mathbb{R}^{1,1}$ or dS$_2$ with $\Sigma_{\gamma,2k-2}$ (considering $\gamma = 1,0,-1$ accordingly). Note that $\tilde{S}$ identically vanishes for these spacetimes, and therefore, once $\gamma$ is fixed, we consider it as the reference background. The mass is computed by evaluating the Noether charge associated to the timelike Killing vector $\partial_t$. Note that the symplectic potential $\Theta$ vanishes for the variation $\delta b=0$, therefore the Noether charge is conserved. The mass of the black hole is then given by

\begin{equation}
    M_k = \mathcal{Q}_{k}(\partial_t)=-\dfrac{2\mu_{2k}}{b} \tilde{S}_{\infty}^{k-1}[k(b^2-4)+2(k-1)\tilde{S}_{\infty}]\,\textrm{Vol}(\Sigma_{\gamma,2k-2})\,,
\end{equation}
where $\displaystyle \tilde{S}_{\infty} = \lim_{r\to\infty}\tilde{S}(r) = 2(\gamma-c)$. Therefore, we obtain
\begin{equation}
    \label{mass-k-dimensions}
    M_k =-\frac{2^{k}\mu_{2k}}{b}\,(\gamma-c)^{k-1}
        \bigl[k\,b^{2}-4(k-1)c-4\gamma \bigr]\,
        \textrm{Vol}\bigl(\Sigma_{\gamma, 2k-2}\bigr)\,,
\end{equation}
and the expressions for dimensions $D=4,6,$ and $8$ are respectively given by
\begin{align}
\label{wald-mass-k2} M_2  &= \frac{4 (c-\gamma)}{3b}\left(b^2-2c-2\gamma\right)\, \textrm{Vol}(\Sigma_{\gamma,2})\\[1ex]
M_3  &= -\frac{39 (c-\gamma)^2}{50b}\left(3b^2-8c-4\gamma\right)\, \textrm{Vol}(\Sigma_{\gamma,4})\\[1ex]
\label{wald-mass-k4} M_4 &= \frac{25960 (c-\gamma)^3}{3087b}
\left(b^2-3c-\gamma\right)\, \textrm{Vol}(\Sigma_{\gamma,6})\,.
\end{align}

The behavior of the black hole free energy as a function of temperature for the solution \eqref{newf} is shown in Fig.~\ref{fig:FTplotk3} and Fig.~ \ref{fig:FTplotk4}. As can be seen, similar to the four-dimensional black hole discussed in the previous section, the system exhibits a zeroth-order phase transition.

\begin{figure}
    \centering
    \begin{tikzpicture}

        \node[anchor=south west] (main) at (0,0)
        {\includegraphics[width=0.6\textwidth]{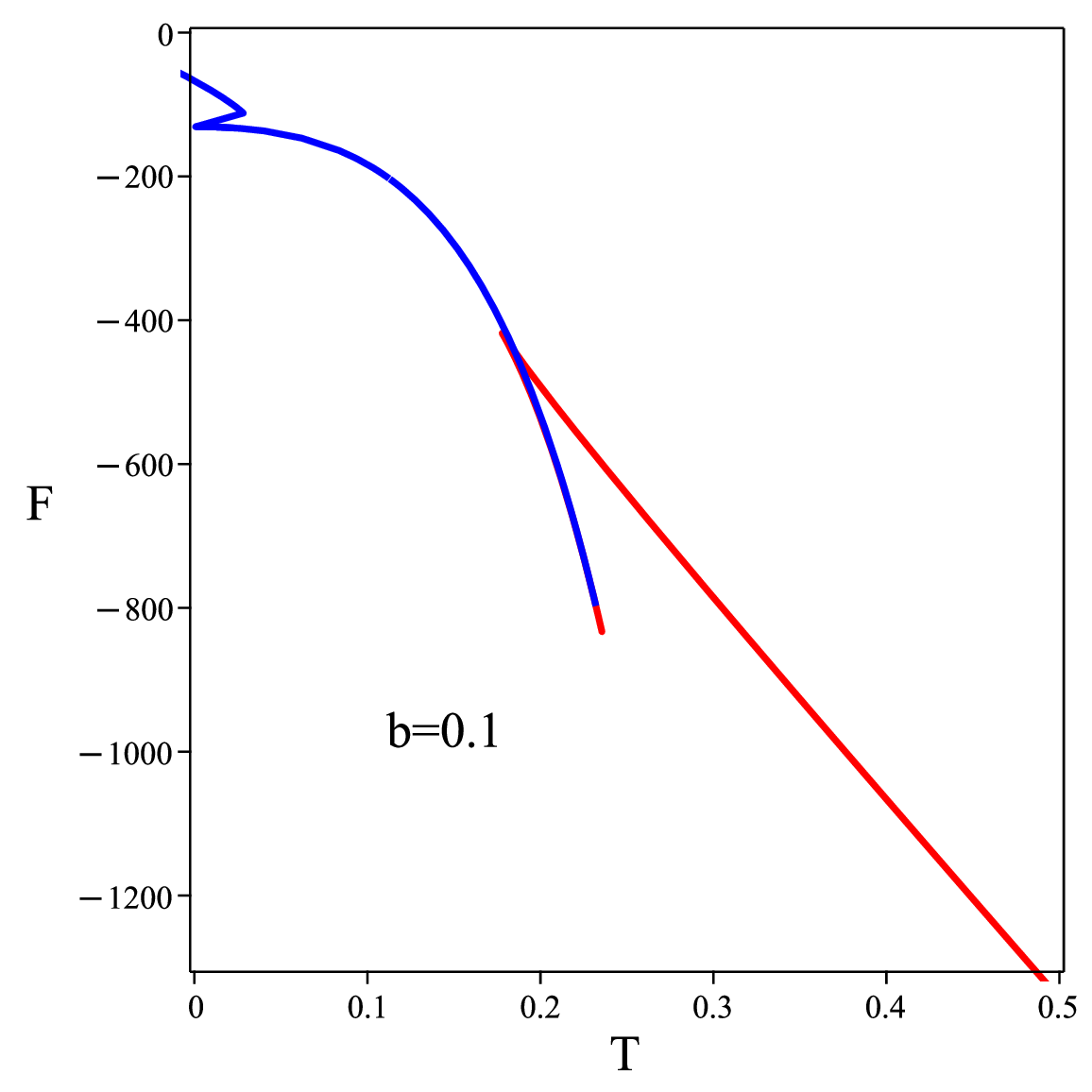}};

        \node[
            anchor = south west,
            draw=black,
            inner sep=1pt,
            fill=white
        ] (inset) at (7.43cm,5.87cm)
        {\includegraphics[width=0.33\textwidth]{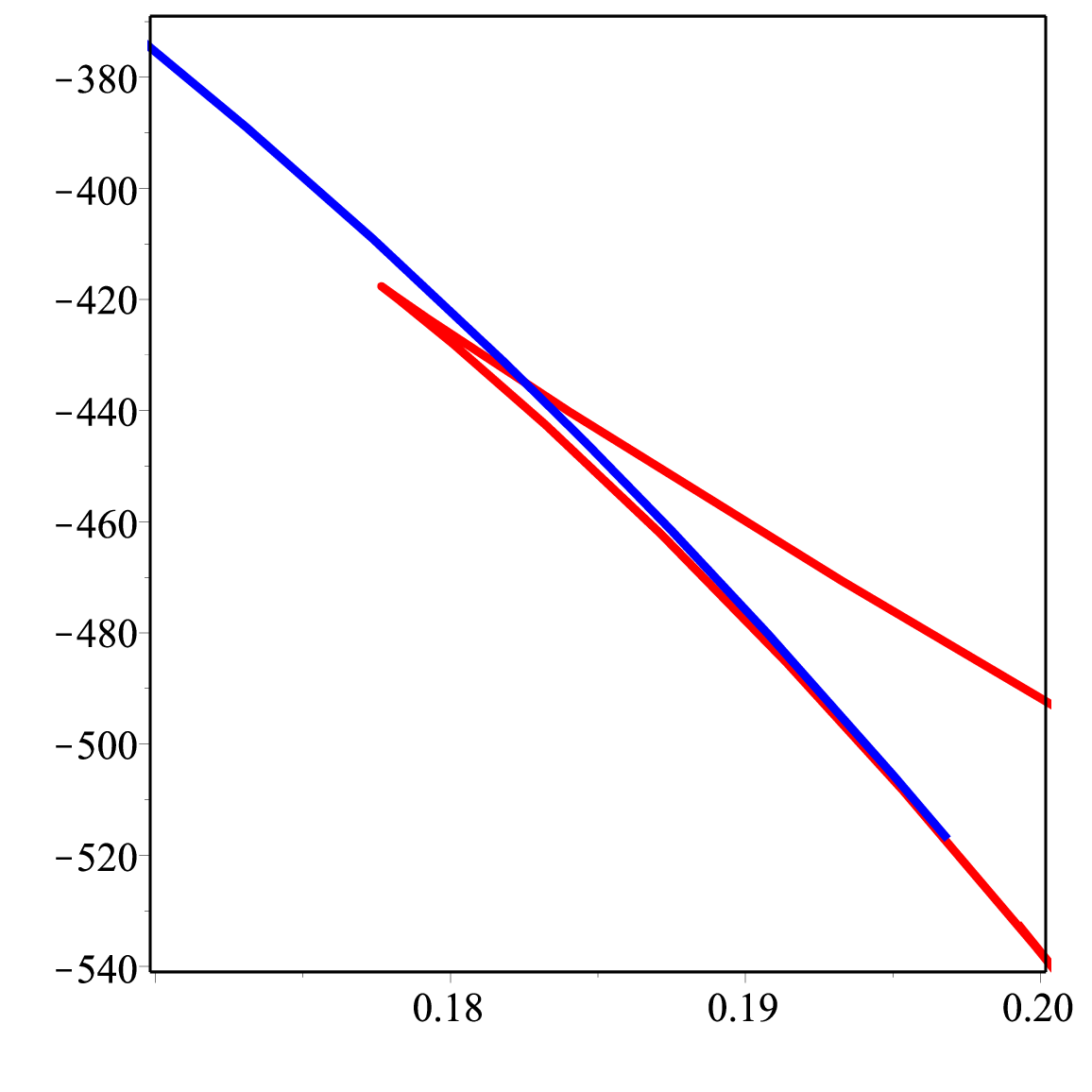}};

        \draw[black, thick, anchor=south west]
        (4.4cm,6.5cm) rectangle (5.4cm,7.2cm);

        \draw[gray, thick, dashed, anchor=south west] 
        (5.4cm,7.2cm) -- (7.43cm,10.9cm);

        \draw[gray, thick, dashed, anchor=south west] 
        (5.4cm,6.5cm) -- (7.43cm,5.87cm);

    \end{tikzpicture}
    \caption{Free energy ${F}$ as a function of temperature ${T}$ for $k=3$ and $\gamma=1$ of \eqref{newf}. The blue and red curves correspond to the small and large black hole branches, respectively. The disconnected branches exhibit a finite jump in the free energy, indicating a zeroth-order phase transition between the small and large black holes. }
    \label{fig:FTplotk3}
\end{figure}
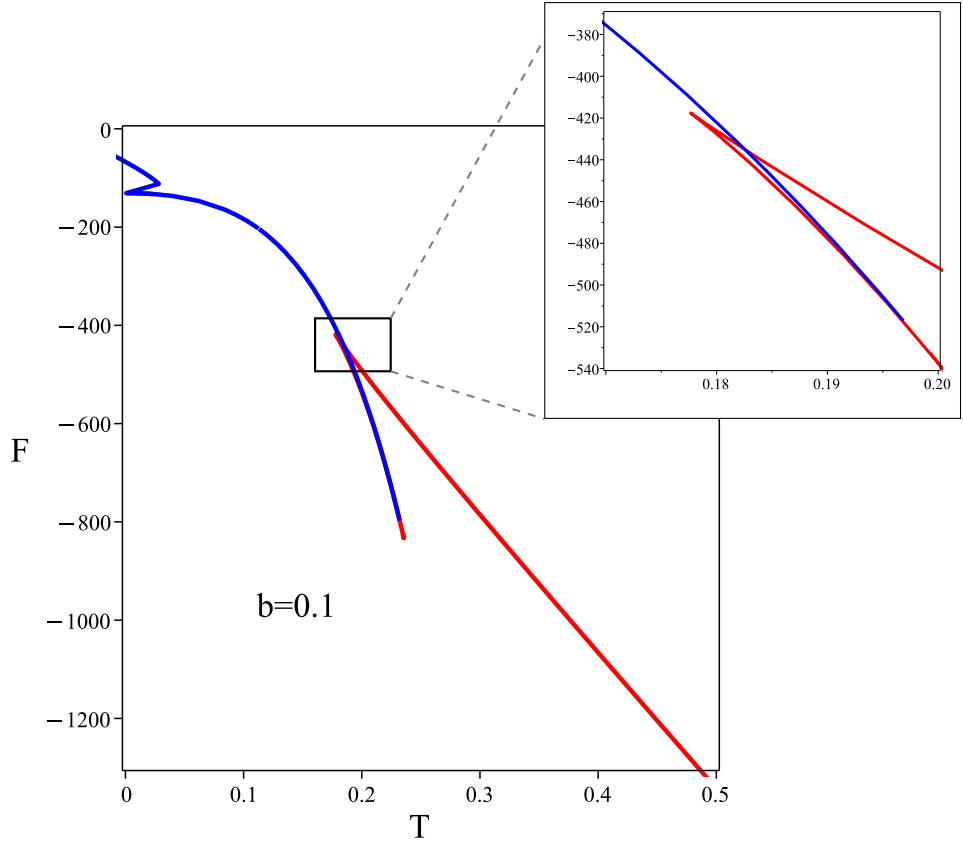
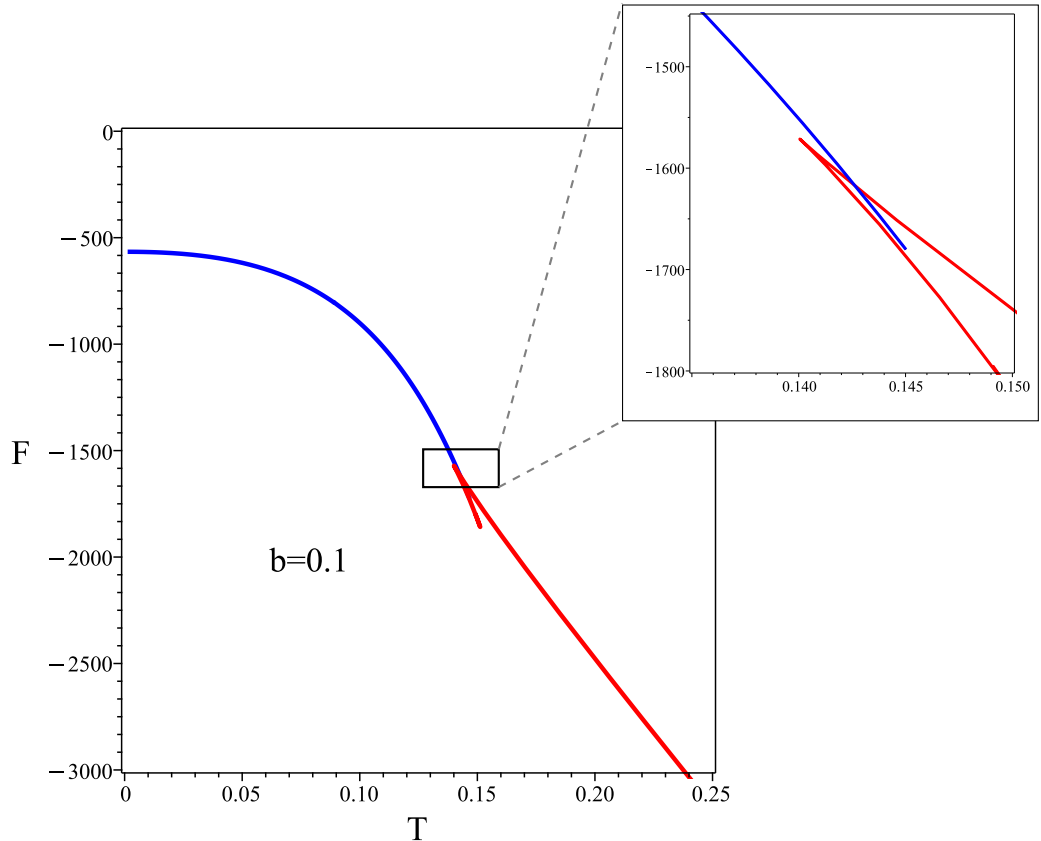
\begin{figure}
    \centering
    \begin{tikzpicture}

        \node[anchor=south west] (main) at (0,0)
        {\includegraphics[width=0.6\textwidth]{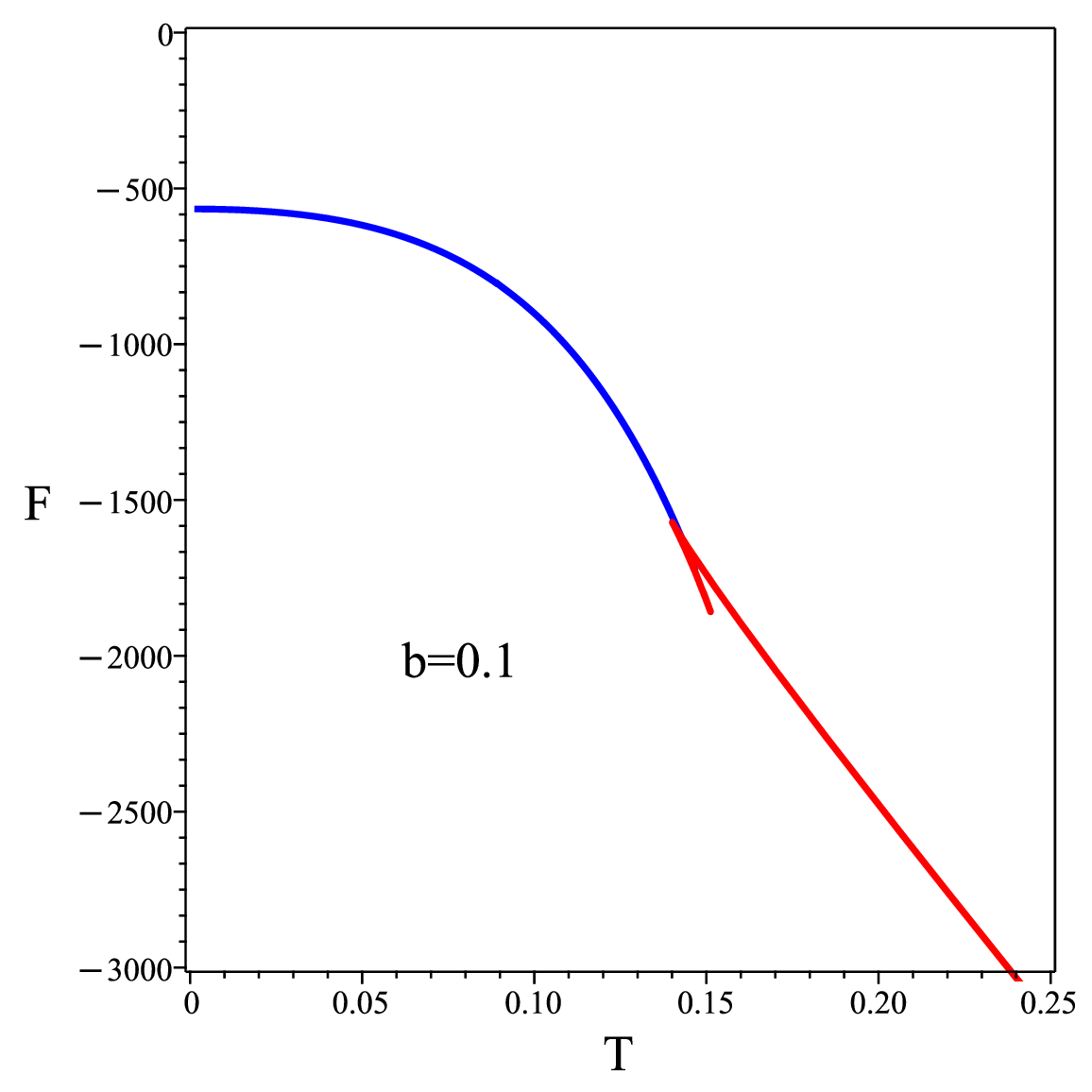}};

        \node[
            anchor = south west,
            draw=black,
            inner sep=1pt,
            fill=white
        ] (inset) at (8.43cm,5.87cm)
        {\includegraphics[width=0.33\textwidth]{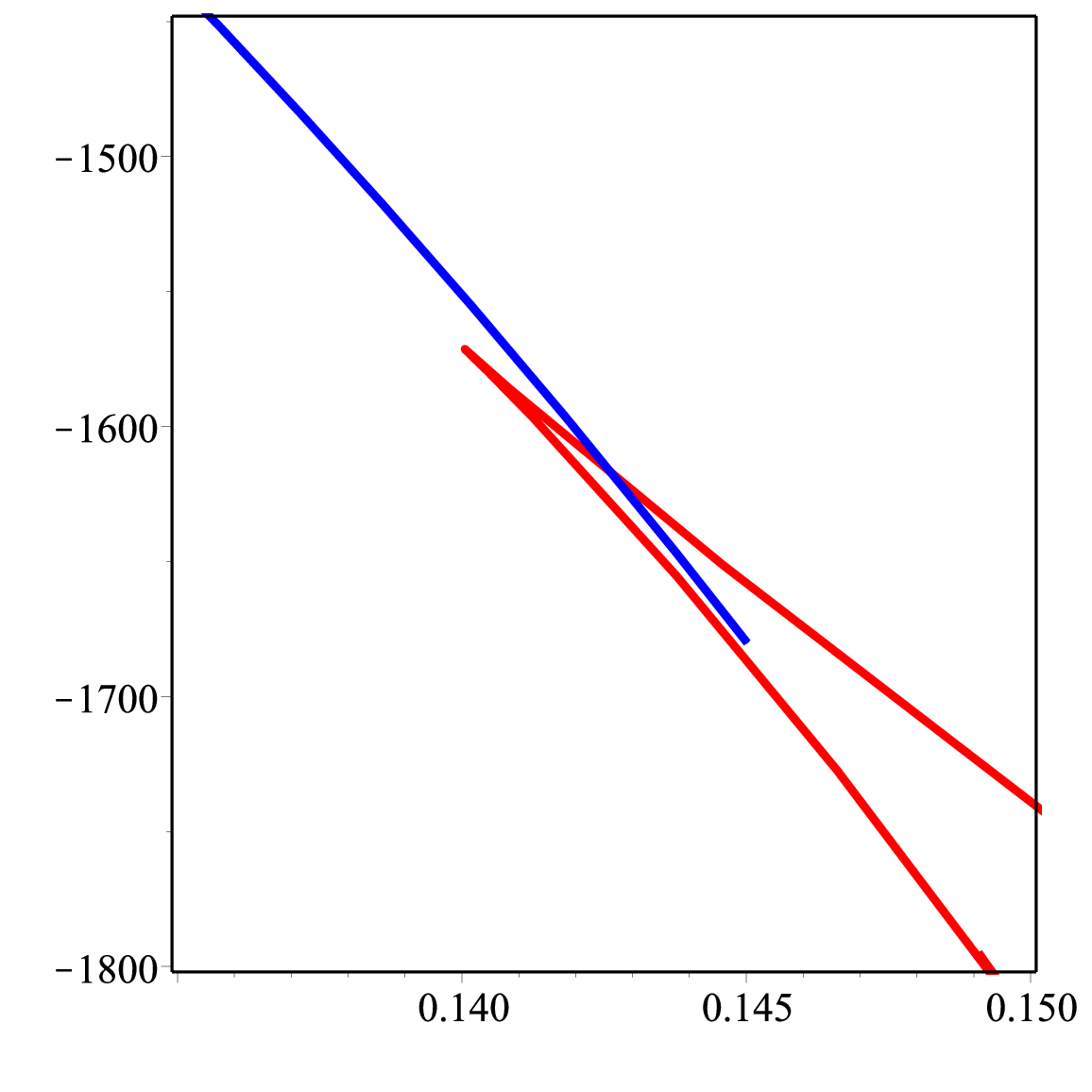}};

        \draw[black, thick, anchor=south west]
        (5.8cm,5cm) rectangle (6.8cm,5.5cm);

        \draw[gray, thick, dashed, anchor=south west] 
        (6.8cm,5.5cm) -- (8.43cm,11.4cm);

        \draw[gray, thick, dashed, anchor=south west] 
        (6.8cm,5cm) -- (8.43cm,5.87cm);

    \end{tikzpicture}
    \caption{Free energy ${F}$ as a function of temperature ${T}$ for $k=4$ and $\gamma=1$ of \eqref{newf}.}
    \label{fig:FTplotk4}
\end{figure}

\section{Conclusion}\label{seccon}
We have addressed the thermodynamic properties of static, spherically symmetric black holes in conformal gravity in arbitrary, even dimensions. These black holes emerge as the result of a Birkhoff's theorem which states that, up to the local conformal redundancies of the theory, spherical symmetry implies the existence of an extra Killing vector, which is timelike in the exterior region of a black hole. We have focused on asymptotically AdS black holes, which are characterized by three integration constants. One of the integration constants corresponds to the curvature radius of the asymptotically locally AdS region, while a second constant determines the strength of a term that is linear in $r$, which is universal across all dimensions. We consider boundary conditions such that both of these quantities are held fixed, and non-vanishing. The remaining integration constant determines the radius of the black hole horizon, and controls a universal $r^{-1}$ terms that appear in the asymptotic expansion of the lapse function. Under these boundary conditions, the charges computed with the Covariant Phase Space approach become integrable. Remarkably, we found that for every dimension, the asymptotic behavior is universally given by \eqref{universalasymp}. With the thermodynamic quantities of the black holes, we have unveileded a new family of zeroth order phase transitions, at a given temperature, between small and large black holes. This phase transition is present in every dimension.

Given the recent interest in conformal gravity, it would be interesting to further explore these theories. For example, it would be interesting to see whether in spite of the six-order nature of the conformal gravity theory introduced by Lü, Pang and Pope in \cite{Lu:2011ks,Lu:2013hx}, such theory still fulfills a Birkhoff's theorem, and whether similar phase transitions as those discovered in this work are present in such higher-derivative framework, in dimension six and even in dimension eight in the recently discovered eighth-order conformal gravity whose Lagrangian contains terms of the form $\mathcal{R}\square^2\mathcal{R}+\ldots$, by Boulanger and Rovere in \cite{Boulanger:2025oli}. It would also be interesting to extend the results of our work, and generalize the family of wormholes in four-dimensional conformal gravity in \cite{Hohmann:2018shl}. We expect to report on some of these points in the future.

\section*{Acknowledgements}
We thank Petarpa Boonserm, Cristóbal Corral, Gastón Giribet and Marcelo Oyarzo for enlightening comments. This work is supported in part by the FONDECYT grants 1230853, 1242043, 1250133, 1262452, 126241 and 11260910. This work is also supported by the Second Century Fund (C2F).

\end{document}